\documentclass[12pt,a4paper]{article}

\usepackage{amsmath}
\usepackage{graphicx}
\usepackage{amsmath}
\usepackage{amsthm}
\usepackage{subcaption}
\usepackage{multirow}
\usepackage{amscd}
\usepackage[T1]{fontenc}

\newtheorem{theorem}{Theorem}

\newtheorem{corollary}[theorem]{Corollary}

\newtheorem{proposition}[theorem]{Proposition}

\title{Rendezvous on the Line with Different Speeds and Markers that can be Dropped at Chosen Time.}

\author{Pierre leone, Nathan Cohen}

\begin{document}
\maketitle

\begin{abstract}
In this paper we introduce a Linear Program (LP) based formulation of a Rendezvous game with markers on the infinite line and solve it. In this game one player moves at unit speed while the second player moves at a speed bounded by $v_{max}\le 1$. We observe that in this setting a slow moving player may have interest to rest still instead of moving. This shows that in some conditions the wait-for-mummy strategy is optimal. We observe as well that the strategies are completely different if the player that holds the marker is the fast or slow one. Interestingly, the marker is not useful when the player without marker moves slowly, i.e. the fast moving player holds the marker.

\end{abstract}

\section{Introduction}
In this article we introduce variation of the asymmetric Rendezvous problem on the line that was  introduced by Alpern and Gal \cite{alperngal1995}. In the original setup, two players are placed on a line at a known distance $D$ and move on the line to Rendezvous. The player's  strategies may be different and start at the same time and both players move at the same speed $v=1$. At the start and while moving the players look in a fixed direction, right or left, say. The directions are chosen randomly each with probability $1/2$. It results that players move either in the direction they look to or the other one, i.e. Forward or Backward. A strategy is a succession of Forward and Backward moves. The optimal solution of this problem is shown to be $13D/8$.
Many variations have been proposed showing that even a simple topology as the infinite line leads to interesting problems. Among the hardest seems to be the symmetric Rendezvous on the line where the two players have to play the same strategy. Partial solutions of this problem are obtained. In \cite{doi:10.1137/S0363012993249195} a strategy is proposed that ensures the Rendezvous time satisfies $R\le5 D$. Subsequent strategies are proposed in \cite{andersonessegaier1995} and \cite{baston1999}  that use the same technique as in \cite{doi:10.1137/S0363012993249195} and reduce the Rendezvous time to $R\le 2.28338 D$ and $R\le 2.2091 D$ respectively. \cite{Uthaisombut2005SymmetricRS} generalizes the technique and improves the bound to $R\le 2.19653 D$. The best known bound $R\le 2.1287 D$  is given in \cite{han2008}.  

Some papers deal with the problem where the initial distance between the players is unknown, see for instance \cite{bastongal1998,alpernbeck1999,alpernbeck2000,ozsoyeller2013}. Usually, the distance is characterized by a probability function. Note that if the players use strategies tailored for a known distance $D$ with Rendezvous time $R$  to a problem where it is only known that the distance is bounded by $D$ then the Rendezvous time for this problem  is bounded by $R$. 

The way time enters the game leads to relevant variations as well. The constraint that the players start at the same time may be relaxed and this leads to asynchronous Rendezvous problems \cite{stachowiak2009asynchronous,czyzowicz2018linear}. Asynchronous Rendezvous problems may assume that an adversary chooses the starting delay or the clocks are assumed to drift with different speeds. There are relations between problems where clocks drift at different speeds and ones where players move at different speed \cite{czyzowicz2018linear}.

A problem where players move on the line and share similarities with the Rendezvous problem is the Group Search Problem on the Line \cite{chrobak2015group}. This problem is motivated by the evacuation problem where players must simultaneously gather at some point. One may imagine that people need to leave a building and are helped with a line drawn on the floor but do not know the right direction to follow. When players move at different speeds, interesting strategies can be found where a fast player move to help slower players.  

Problems where players move on a circle share similarities with problems on the line \cite{howard1999,kranakis2003mobile,flocchini2004multiple,kranakis2010mobile,di2017optimal}. Compare to the infinite line the ring is a compact topology but symmetry breaking has to be solved as well to ensure Rendezvous. Tokens may be left by players \cite{czyzowicz2008power,flocchini2004mobile,das2008rendezvous}. For Rendezvous problems on the line \cite{baston2001rendezvous,leone2018rendezvous} present results where markers are used by players. With a more robotic and computer science flavor some problems encompass faulty agents \cite{das2015mobile,das2019gathering}.

Rendezvous problems are far from being limited to the infinite line or ring. Problems may be stated for agents moving on the plane, on graphs, on a torus, on networks and so on, see for instance \cite{alpern2006theory,alpern2013ten,pelc2019deterministic}. These problems are different from the ones considered in this paper.

Markers can have different effects. For instance, the game may end at the time the marker is found, i.e. Rendezvous occurs or the marker is found. This would be the case if a phone number is written on the marker. With such a marker, the game would be close to a version of search-and-rescue game \cite{DBLP:journals/eor/Lidbetter20}. This may be seen as a mix of Rendezvous and Search games, see for instance \cite{bastonkikuta,baston2019search,doi:10.1002/net.21504} for search games on graphs and \cite{hohzaki2016search,alpern2013search,alpern2006theory} for general references.

\section{Our contributions}
In this paper we consider the (synchronous) Rendezvous problem on the line with known initial distance $D$ where players move at different speeds and where a marker can be left by one of the players. Without loss of the generality we assume that one player moves at speed $1$ while the second player moves at speed $v\le 1$. We show that investigations can be conducted with linear programming techniques to identify optimal strategies. This is not the conventional approaches in the literature where the results are usually guessed and optimality is subsequently proved.  The reduction of rendezvous search game to another formalism to be solved appears in the literature, see for instance \cite{alpern1999asymmetric}. Here, the reduction to parametric linear programming has the further advantage that the same method can be applied to compute different measures of optimality. For instance, the optimization of the last rendezvous time. Actually, any linear combination of the rendezvous times can be optimized.

In \cite{leonealpernNRL2018} a similar problem with markers is considered by one of the author of the present article. However, the techniques of proof are completely different. The parametric linear programming approach of this article leads to more precise results than the ad-hoc approach of \cite{leonealpernNRL2018}. Moreover, here we accommodate to players with different speeds.

\section{Problem formulation}
We begin by presenting the formalization of the problem  as given in \cite{alperngal1995}. Two players, $I$ and $II,$
are placed at distance $D=1$\footnote{However, the results depend linearly on the initial distance and are stated for general $D$.}  apart on the real line, and faced in random directions which they call ``Forward''. Their common aim is to minimize the expected amount of time required to meet. They each know the distance $1$ but not the direction the other player is facing. It is not a restriction to assume that player I's starting point is located at position $0$ of the line and his speed is bounded by $v \le 1$. His position is given by a function $f\left( t\right)
\in \mathcal{F(\alpha)}$ where 
\begin{equation}\label{equ:bigspace}
\mathcal{F(\alpha)}~=\left\{ f:\left[ 0,T\right] \rightarrow R,~f\left( 0\right)
=0,\left\vert f\left( t\right) -f\left( t^{\prime }\right) \right\vert \leq
\alpha \left\vert t-t^{\prime }\right\vert \right\} ,
\end{equation}
for some $T$ sufficiently large so that Rendezvous will have taken place.

What are unknown  are the initial position of player II that may be $\pm 1$ and the Forward direction of player II that may point to the positive or negative side of the infinite line. Again without restriction of the generality we assume that the speed of player II is bounded by $1$. Hence, depending on the initial conditions of player II his position at time $t$ is given by $\pm 1 \pm g(t)$ with $g\in \mathcal{F}(1) = \mathcal{F}$.

The Rendezvous times are defined by
\begin{equation}\label{t^1} t^1 = min \{ t : f(t) = 1 + g(t) \}. \end{equation}
when player II is originally located at $+D$ and his Forward direction points the positive side of the line.
\begin{equation}\label{t^2} t^2 = min \{ t : f(t) = 1 - g(t) \}. \end{equation}
when player II is originally located at $+D$ and his Forward direction points the negative side of the line.
\begin{equation}\label{t^3} t^3 = min \{ t : f(t) = -1 + g(t) \}. \end{equation}
when player II is originally located at $-D$ and his Forward direction points the positive side of the line.
\begin{equation}\label{t^4} t^4 = min \{ t : f(t) = -1 - g(t) \}. \end{equation}
when player II is originally located at $-1$ and his Forward direction points the negative side of the line.

It is common in the literature to speak of $4$ agents (of player II) located at positions $\pm 1$  and with Forward direction $\pm 1$ and moving concurrently. Player I need to Rendezvous with the four agents to end the game \cite{alpern2006theory}. Concretely, 
\begin{itemize}
\item agent $1$ is located at $+1$ with Forward direction $+1$ and its Rendezvous time is $t^1$,
\item agent $2$ is located at $+1$ with Forward direction $-1$ and its Rendezvous time is $t^2$, 
\item agent $3$ is located at $-1$ with Forward direction $+1$ and its Rendezvous time is $t^3$,
\item agent $4$ is located at $-1$ with Forward direction $-1$ and its Rendezvous time is $t^4$.
\end{itemize}
The notation $t_1 \le t_2 \le t_3 \le t_4$ denote the Rendezvous times in the order they occur and $(o_i,b_i)$  denote the agent with origin $o_i=\pm 1$ and Forward direction $b_i=\pm 1$. The order of the Rendezvous times is given by the index $i$, $t_1$ is Rendezvous with agent $(o_1,b_1)$, $t_2$ is Rendezvous with agent $(o_2,b_2)$, $t_3$ is Rendezvous with agent $(o_3,b_3)$, $t_4$ is Rendezvous with agent $(o_4,b_4)$. When necessary we use the convention $t_0=0$

The Rendezvous value $R(f,g)$ is defined to be the average value 
\[ R(f,g) = \frac{1}{4}\biggl(t^1+t^2+t^3+t^4\biggr).\]

Finally, the Rendezvous value of the game is defined by
\begin{equation}\label{eq:RDVvalue}
 R = \min\biggl\{R(f,g) : f\in \mathcal{F}(v), g\in \mathcal{F}\biggr\}.
 \end{equation}

A first remark that simplifies the problem is that the functional spaces $\mathcal{F}(v), \mathcal{F}$ may be reduced to consider only functions $f\in\mathcal{F}(v), g\in\mathcal{F}$ {\it whose speed is constant between the Rendezvous times}. Indeed, if the speed is not constant, moving at the average speed between Rendezvous times leads to the same Rendezvous value. Moreover, similarly to Lemma 5.1  of \cite{alperngal1995} or Theorem 16.10 of \cite{alpern2006theory} or Proposition 3 of \cite{leone2018rendezvous} we have the following result for $g\in\mathcal{F}$.
\begin{proposition}\label{prop:optstrat} If $v\le 1$ then for the optimal strategies the function $g\in\mathcal{F}$ is of constant slope equal at $\pm 1$ between the Rendezvous, i.e. the speed of the fast player is always maximal.
\end{proposition}
\begin{proof} Let us assume that player II whose position is given by function $g\in\mathcal{F}$ and initial position does not move at maximal speed between Rendezvous times $t_{i-1}< t_i$. This means that player II can reach the Rendezvous position at a time $t_i-\epsilon$ with $\epsilon>0$. By moving faster it may happen that player II Rendezvous with player I before time $t_i$ reducing the Rendezvous time $t_i$. If not we modify the trajectory of player II in the following way. After reaching the Rendezvous point at time $t_i-\epsilon$ player II continues in the same direction for a period $\epsilon/2$ and then goes the other way for a period $\epsilon/2$ back to the Rendezvous position at time $t_i$. At time $t-\epsilon/2$ player I must be at a distance less than $v\epsilon/2 $ to the Rendezvous position and because player II is at a distance $\epsilon/2$ and $\epsilon/2\ge v\epsilon/2 $ the Rendezvous must occurs before time $t_i$.
To summarize, by moving at full speed player II always reduces the Rendezvous time $t_i$. After time $t_i$ player II follows the original strategy and the remaining Rendezvous times are not changed. In total the modified strategy reduces the Rendezvous value showing that the original strategy is not optimal.
We emphasize the the fast moving player moves at maximal speed while the slow moving player can move at any speed in $[0,v]$.
\end{proof}
\begin{corollary}\label{cor:optstrat} We assume here that the speed of player I is bounded by $v\le 1$ and the speed of player II by $1$, i.e. $f\in\mathcal{ F}(v)$, $g\in\mathcal{F}$. The sets of optimal strategies $(f,g)$ for player I and II respectively are given by
\begin{align}
f(t)=\begin{cases} v_1\cdot t, & t\in [0,t_1]\\ v_1\cdot t_1+ v_2\cdot (t-t_1),&t\in[t_1,t_2]\\ v_1\cdot t_1+ v_2\cdot (t_2-t_1)+v_3\cdot (t-t_2),&t\in[t_2,t_3]\\v_1\cdot t_1+ v_2\cdot (t_2-t_1)+v_3\cdot (t_3-t_2)+v_4\cdot (t-t_3),&t\in[t_3,t_4]\end{cases}\label{eq:fnomarker} \\
g(t)=\begin{cases} d_1\cdot t,&t\in [0,t_1]\\ d_1\cdot t_1+ d_2\cdot (t-t_1),&t\in[t_1,t_2]\\ d_1\cdot t_1+ d_2\cdot (t_2-t_1)+d_3\cdot (t-t_2),&t\in[t_2,t_3]\\d_1\cdot t_1+ d_2\cdot (t_2-t_1)+d_3\cdot (t_3-t_2)+d_4\cdot (t-t_3),&t\in[t_3,t_4]\end{cases} \label{eq:gopt}
\end{align}
with $v_i\in[-v,v]$ and $d_i=\pm 1$ and $t_1\le t_2\le t_3\le t_4$ are the Rendezvous times.
\end{corollary}
Proposition \ref{prop:optstrat} and Corollary \ref{cor:optstrat} are not new and constantly used in the literature, see for instance Chapter 17.1 of \cite{alpern2006theory}. We stress that player I having the smallest speed bound may move at a slower speed than the maximal one. Indeed, we will observe that for $v$ small optimal strategy for player I is to not move before $t_2$. The ``wait for mummy''  strategy is then  optimal for starting the game.

We consider that player I has at disposal a marker  that may be left at a chosen time. The marker helps player II that stops following the strategy after finding the marker and continues in the same direction at maximal speed until Rendezvousing with player I. The same arguments as the ones in Proposition  \ref{prop:optstrat} and Corollary \ref{cor:optstrat} or Proposition 3 of \cite{leone2018rendezvous} show that player I move at constant velocity before and after dropping the marker. There are four different cases to consider for the formulation of the problem depending on which interval, $[0,t_1], [t_1,t_2], [t_2,t_3], [t_3,t_4]$ player I drops off the marker. This leads to the following Proposition that characterizes the optimal strategies.

\begin{corollary}\label{cor:optstratv} When player I has a marker that can be dropped off at chosen time $z$, the set of optimal strategies $f$ for player I are given by
\begin{equation}\label{eq:fbeforet_1}
f(t)=\begin{cases} v_0\cdot t, &t\in [0,z]\\v_0\cdot z+v_1\cdot(t-z), & t\in [z,t_1]\\ v_0\cdot z+v_1\cdot (t_1-z)+ v_2\cdot (t-t_1),&t\in[t_1,t_2]\\ v_0\cdot z+v_1\cdot (t_1-z)+ v_2\cdot (t_2-t_1)+v_3\cdot (t-t_2),&t\in[t_2,t_3]\\v_0\cdot z+v_1\cdot (t_1-z)+ v_2\cdot (t_2-t_1)+v_3\cdot (t_3-t_2)+v_4\cdot (t-t_3),&t\in[t_3,t_4]\end{cases} \\
\end{equation}
if $z\in[0,t_1]$.
\begin{equation}
f(t)=\begin{cases} v_1\cdot t, & t\in [0,t_1]\\ 
			v_1\cdot t_1+v_0\cdot (t-t_1),&t\in[t_1,z]\\ 
			v_1\cdot t_1+v_0\cdot (z-t_1)+v_2\cdot (t-z),&t\in[z,t_2]\\
			v_1\cdot t_1+v_0\cdot (z-t_1)+v_2\cdot (t_2-z)+ v_3\cdot (t-t_2),&t\in[t_2,t_3]\\
			v_1\cdot t_1+v_0\cdot (z-t_1)+v_2\cdot (t_2-z)+ v_3\cdot (t_3-t_2)+v_4\cdot (t-t_3),&t\in[t_3,t_4]\end{cases} \\
\end{equation}
if $z\in[t_1,t_2]$.
\begin{equation}
f(t)=\begin{cases} v_1\cdot t, & t\in [0,t_1]\\ 
			v_1\cdot t_1+v_2\cdot (t-t_1),&t\in[t_1,t_2]\\ 
			v_1\cdot t_1+v_2\cdot (t_2-t_1)+v_0\cdot (t-t_2),&t\in[t_2,z]\\
			v_1\cdot t_1+v_2\cdot (t_2-t_1)+v_0\cdot (z-t_2)+ v_3\cdot (t-z),&t\in[z,t_3]\\
			v_1\cdot t_1+v_2\cdot (t_2-t_1)+v_0\cdot (z-t_2)+ v_3\cdot (t_3-z)+v_4\cdot (t-t_3),&t\in[t_3,t_4]\end{cases} \\
\end{equation}
if $z\in[t_2,t_3]$.
\begin{equation}
f(t)=\begin{cases} v_1\cdot t, & t\in [0,t_1]\\ 
			v_1\cdot t_1+v_2\cdot (t-t_1),&t\in[t_1,t_2]\\ 
			v_1\cdot t_1+v_2\cdot (t_2-t_1)+v_3\cdot (t-t_2),&t\in[t_2,t_3]\\
			v_1\cdot t_1+v_2\cdot (t_2-t_1)+v_3\cdot (t_3-t_2)+ v_0\cdot (t-t_3),&t\in[t_3,z]\\
			v_1\cdot t_1+v_2\cdot (t_2-t_1)+v_3\cdot (t_3-t_2)+ v_0\cdot (z-t_3)+v_4\cdot (t-z),&t\in[z,t_4]\end{cases} \\
\end{equation}
if $z\in[t_3,t_4]$.
The optimal strategies for player II are still of the form of Equation $(\ref{eq:gopt})$. In any cases the parameters are constrained to  $v_i\in[-v,v]$ ($v\le 1$)and $d_i=\pm 1$ and $t_1\le t_2\le t_3\le t_4$ are the Rendezvous times.
\end{corollary}

Corollaries \ref{cor:optstrat} and \ref{cor:optstratv} are very useful in making the Rendezvous value of the game given by Equation \ref{eq:RDVvalue} computable. Indeed, the set of functions $(f,g)$ to be considered is finite. Notice that for the problem with marker the set is finite provided that $v$ is fixed.

It is crucial to point out that in Corollary \ref{cor:optstratv} the optimal strategy of player II is of the form of Equation $(\ref{eq:gopt})$, but if the marker is found at some time player II no longer follows the strategy but continues in the same direction thereafter. For instance, if the marker is found in the interval $[0,t_1]$ at time $t_z$ we must have $z\le t_z$ and the condition
\begin{equation*}
o+d_1 b t_z=v_o z,
\end{equation*}
must be satisfied where $o$ is the original strating point of player II ($o=\pm 1$) and $b$ is the Forward direction of player II ($b=\pm 1$). Indeed, the condition states that player II starting at position $o$ at time $0$ is at the marker's position at time $t_z$, i.e. the marker is found. The coefficient $d_1$ is given by the strategy followed by player II and $d_1\cdot b$ is the effective motion depending on the Forward direction $b$. Thereafter, player II does not follow the strategy but continues in the same direction, i.e. substitutes $d_1$ for $d_i$ in Equation $(\ref{eq:gopt})$. Hence, if Rendezvous does not occur in $[0,t_1]$ player II motion is given by
\begin{equation*}
o+d_1 b t_z + d_1 b (t_1-t_z) = o + d_1 b t_1.
\end{equation*}
If no Rendezvous occurs in the next time interval $[t_1,t_2]$ the motion of player II is given by 
\begin{equation*}
 o + d_1 b t_1 + d_1 b (t_2-t_1),
\end{equation*}
(compare with the second line of Equation $(\ref{eq:gopt})$). The same reasoning applies for the next time intervals $[t_2,t_3]$ and $[t_3,t_4]$.

\section{Solution of the problem without marker}\label{sec:nomarker}
The optimal strategies of the problem without marker are given in Corollary \ref{cor:optstrat}. There are $8$ unknowns $v_i$ and $d_i$, the Rendezvous time being a consequence of these variables. To reduce the problem to a family of linear programs we rewrite the strategies to remove the products $v_i\cdot t$ by introducing new variables $vt_i$ with the bounds $0\le vt_i \le v\cdot(t_i-t_{i-1})$ with $v$ is the maximal speed of player I. To take into account that the speed of player I may be negative (when player I is going Backward) we introduce new variables $a_i=\pm 1$ and the motion is computed relatively to $a_i\cdot vt_i$. Notice that $a_i$ is a parameter which is fixed before calling the LP-solver (hence the problem is still linear). 

To define the meeting times $t_i$ it is needed to specify in which order they occur. Ordering the meeting times amounts to chose a permutation $\sigma$ of $\{1,2,3,4\}$ such that $t_i=t^{\sigma(i)}$ where $t^i$ are defined by equations $(\ref{t^1})$,$(\ref{t^2})$,$(\ref{t^3})$,$(\ref{t^4})$. For this, we introduce new variables $(o_i,b_i)$ with $o_i=\pm 1$ and $b_1=\pm 1$ to refer to specific agents of player II. Concretely,
\begin{itemize}
\item agent $1$ is referred by $(o_i=+1, b_i = +1)$ and the Rendezvous time $t^1$ is defined by $(\ref{t^1})$,
\item agent $2$ is referred by $(o_i=+1, b_i = -1)$ and the Rendezvous time $t^2$ is defined by $(\ref{t^2})$, 
\item agent $3$ is referred by $(o_i=-1, b_i = +1)$ and the Rendezvous time $t^3$ is defined by $(\ref{t^3})$, 
\item agent $4$ is referred by $(o_i=-1, b_i = -1)$ and the Rendezvous time $t^4$ is defined by $(\ref{t^4})$.
\end{itemize}
Rendezvous always occur in the order $(o_1,b_1)$, $(o_2,b_2)$, $(o_3,b_3)$, $(o_4,b_4)$, The values of $o_i$ and $b_i$  

\begin{align}
&\underset{\Delta _{i}}{\text{min}}~~t _{1}+t _{2} +t_{3}+t _{4}  \notag \\
& o_1 + d_1 b_1 t _1 = a_1 vt_1 \notag \\
& o_2 + d_1 b_2 t_1 + d_2 b_2(t_2-t_1) = a_1 vt_1+a_2 vt_2 \notag \\
& o_3 + d_1 b_3 t_1 + d_2 b_3 (t_2-t_1) + d_3 b_3 (t_3-t_2) = a_1 vt_1+a_2 vt_2 + a_3 vt_3 \notag \\
& o_4 + d_1 b_4 t_1 + d_2 b_4 (t_2-t_1) + d_3 b_4 (t_3-t_2) + d_4 b_4 (t_4-t_3)= a_1 vt_1+ \notag \\
&~~~~~~~~~~~~~~~~~~~~~~~~~~~~~~~~~~~~~~~~~~~~~~~~~~~~~~~~~~~~a_2 vt_2 + a_3 vt_3 + a_4 vt_4  \notag \\
& 0\le vt_1 \le v \cdot t_1\notag\\
& 0\le vt_2 \le v\cdot  (t_2-t_1)\notag\\
& 0\le vt_3 \le v\cdot  (t_3-t_2)\notag\\
& 0\le vt_4 \le v\cdot  (t_4-t_3)\notag\\ \notag \\
& a_i,b_i, o_i, d_i\in \{0,1\} \notag \sum o_{i}=0, \sum d_{i}=0, ~o_i=o_j \Rightarrow d_i\not = d_j. \notag\\
& t_1 \ge 0, t_2 \ge 0, t_3 \ge 0, t_4 \ge 0\notag\\ \notag
\end{align}

For the computation of the solution, we used the variables $o_i, b_i, d_i, a_i$ as parameters. For each set of values we solve the corresponding linear program. The number of linear programs solved is $1536$ which is solved in a few seconds using Python Gurobi library. Notice that by symmetry we fixed $a_1=1$ and $d_1=1$. 

\begin {figure}[h]
\begin{center}
\includegraphics[scale=0.6]{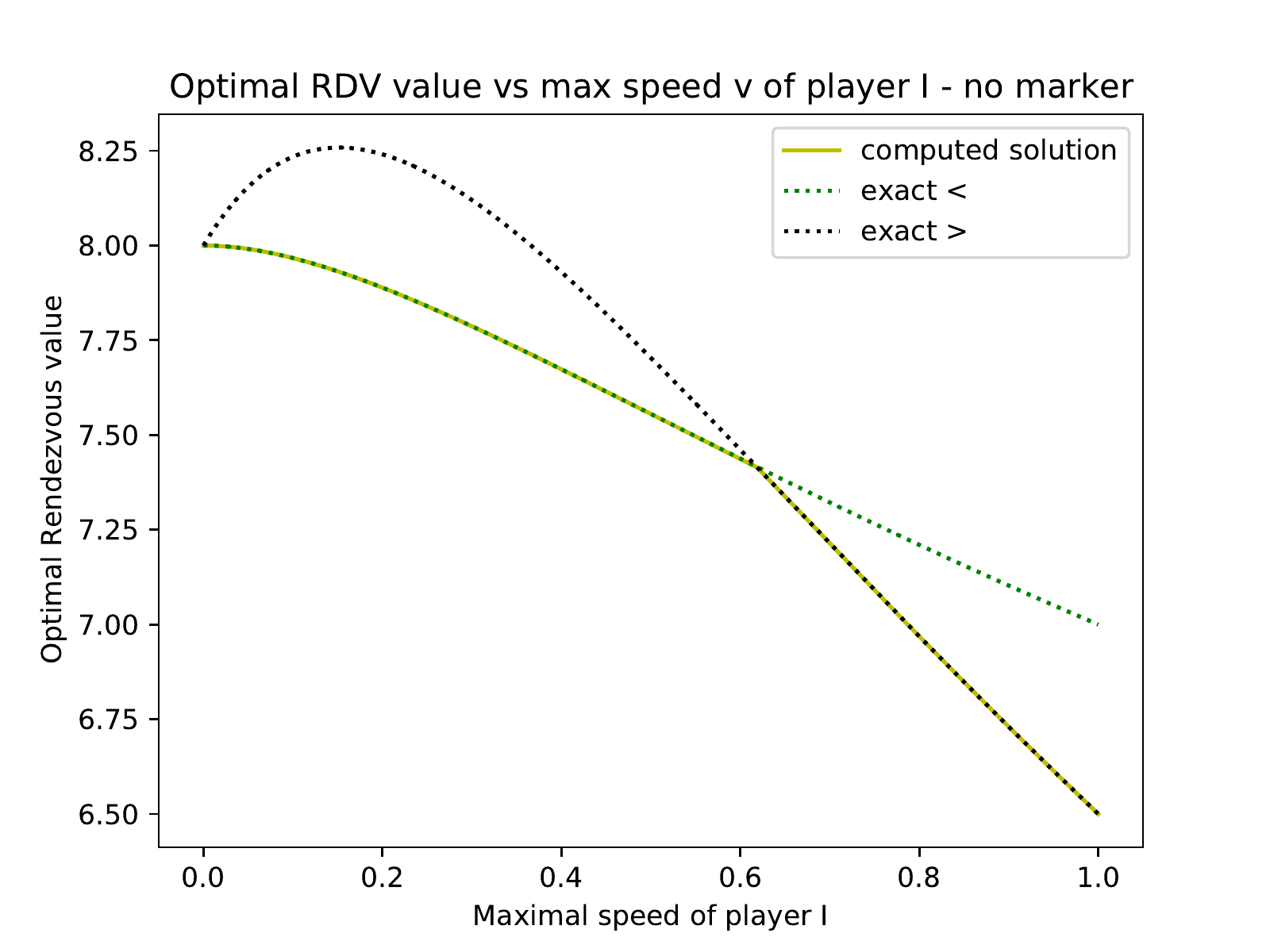}\caption{Plot of the solution of the Rendezvous problem vs the maximum speed of player I. The computed solution is a plot a $1001$ points evaluated at $n/1000$, $n\in[0,1000]$. The exact solution has two algebraic forms for $v<(\sqrt{5}-1)/2$ (exact <, $(\ref{eq:exact<})$) or $v>(\sqrt{5}-1)/2$ (exact >, $(\ref{eq:exact>})$).}\label{fig:optVSv} 
\end{center}
\end{figure}

The plot of the results are shown on Figure \ref{fig:optVSv}, the optimal Rendezvous value is plotted versus the maximal speed of player $I$ for discrete values $n/1000$, $n=0,1,\ldots,1000$. 

\begin {figure}[h]
\includegraphics[scale=0.4]{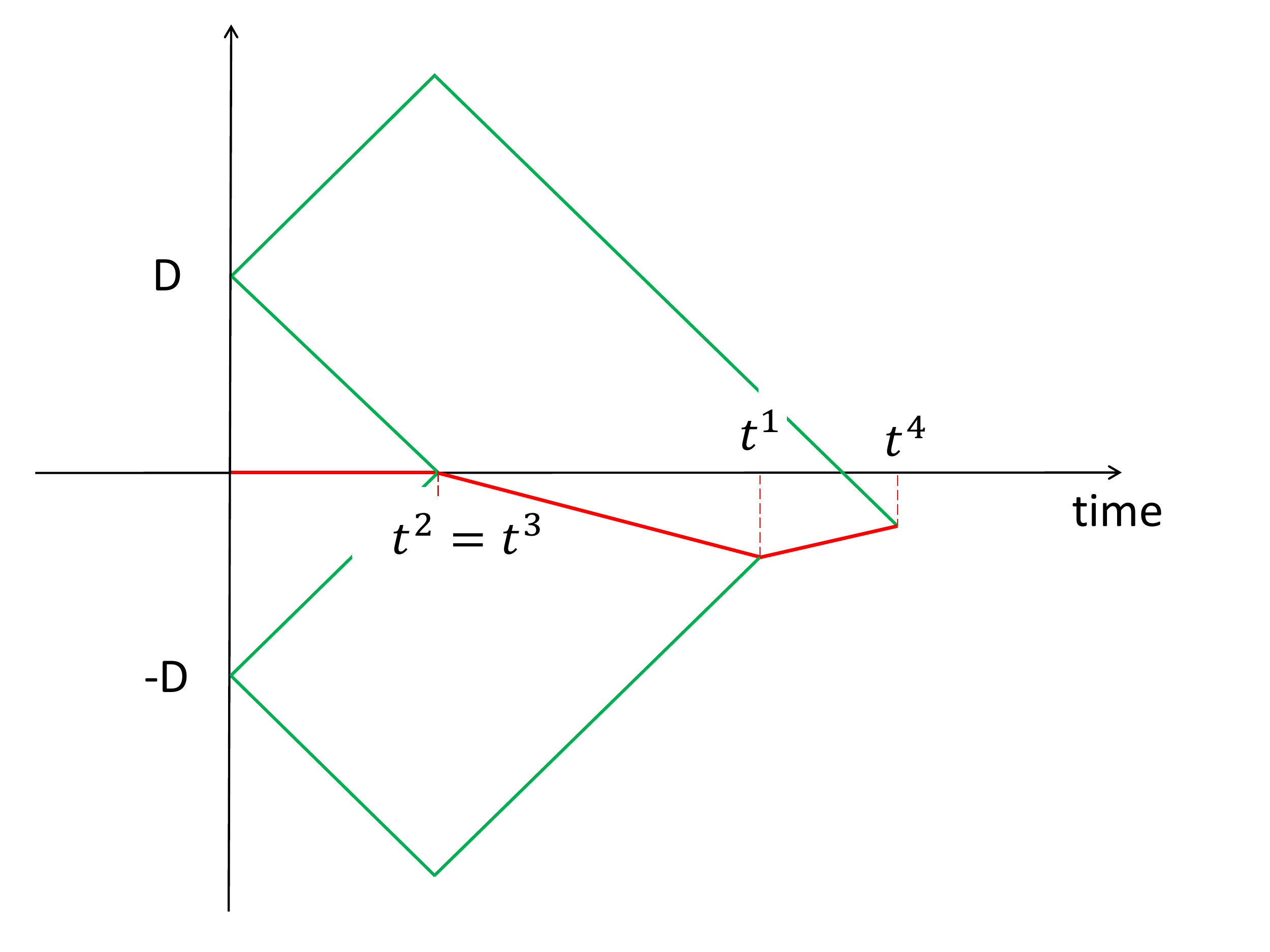}\caption{Optimal strategy for $v\in[0.001,0.618]$.}\label{fig:optVSvpetit} 
\end{figure}

\begin {figure}[h]
\includegraphics[scale=0.4]{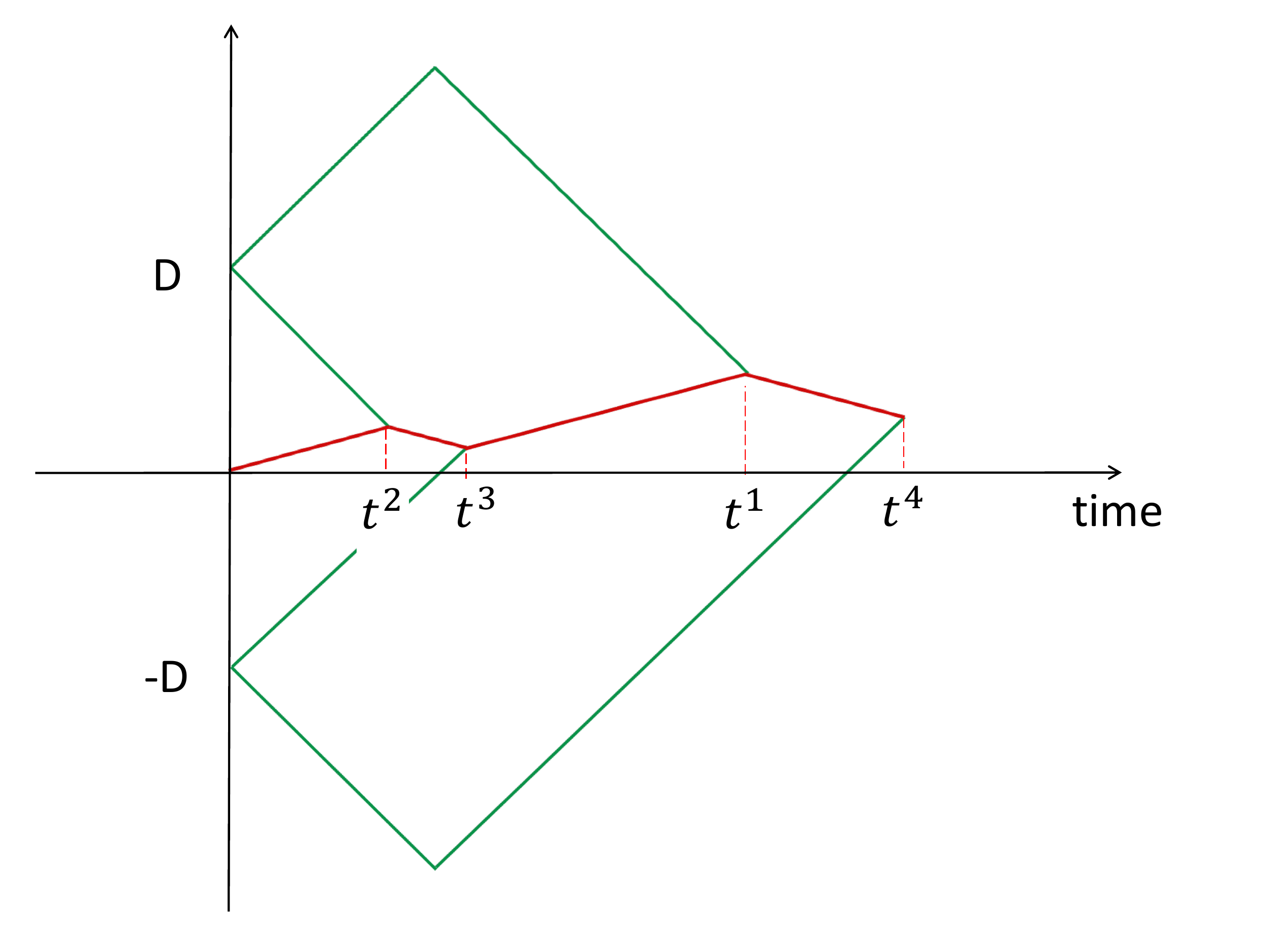}\caption{Optimal strategy for $v\in]0.619, 0.990]$.}\label{fig:optVSvgrand} 
\end{figure}

Besides the computation of the optimal Rendezvous values we record the corresponding optimal strategy. We observe that for $v\le 0.618$ the optimal strategy is $(a_1=0, a_2 =0, a_3=1, a_4=-1)$ for player I and $(d_1=1, d_2 =1, d_3=-1, d_4=-1)$ for player II as illustrated in Figure \ref{fig:optVSvpetit}. In terms of the optimal strategy of Corollary \ref{cor:optstrat} the speeds of player I are $v_1=0$, $v_2=0$, $v_3=-v$, $v_4 = v$. Notice that player I can as well play the symmetric strategy $(a_1=0, a_2 =0, a_3=1, a_4=-1)$ which is not illustrated.

It is relevant to observe that until $t^2=t^3$ the wait for mummy strategy is the optimal strategy for player I, motion starts only after. For this strategy the expressions for the Rendezvous times and Rendezvous value of the game are given by:
\begin{align}
t_1&=1, t_2=1, t_3=\frac{3+v}{1+v}, t_4=\frac{v^2+8v+3}{(1+v)^2},\notag\\ \notag\\
R&=\frac{4v^2+16v+8}{(1+v)^2}. \label{eq:exact<}
\end{align}

Since $v=0.619$ the optimal strategy is a switch to $a_1=1, a_2=-1, a_3=1, a_4=-1$ for player I and $d_1= 1, d_2=1, d_3=-1, d_4=-1$. In terms of the optimal strategy of Corollary \ref{cor:optstrat} the speeds of player I are $v_1=-v$, $v_2=v$, $v_3=-v$, $v_4 = v$. The speed of player I is large enough to allow improving the Rendezvous value by moving from the start. The expressions for the Rendezvous times and Rendezvous value of the game are given by:
\begin{align}
t_1&=\frac{1}{v+1}, t_2=\frac{3v+1}{(v+1)^2}, t_3=\frac{5v+3}{(v+1)^2}, \notag \\
t_4&=\frac{7v^2+14v+3}{(1+v)^3}, R=\frac{16v^2+28v+8}{(1+v)^3}. \label{eq:exact>}
\end{align}

With direct computations we see that $(\ref{eq:exact<})$ is better that $(\ref{eq:exact>})$ for $v\le (\sqrt{5}-1)/2$, see Figure \ref{fig:optVSv}.

It is stated in \cite{alpern2006theory}, Chapter 17.1 that the optimal solution is given by $(\ref{eq:exact<})$ for $v\le (\sqrt{5}-1)/2$ and by $(\ref{eq:exact>})$ for $v\ge (\sqrt{5}-1)/2$. With our linear programming approach we first conclude that $(\ref{eq:exact<})$ is optimal for $v= n/1000\le (\sqrt{5}-1)/2$ and by $(\ref{eq:exact>})$ for $v=n/1000\ge (\sqrt{5}-1)/2$, $n= 0, 1, 2, \ldots$ (discrete values).

However, we can say more. Let us denote $opt(v)$ the function that returns the optimal value of the game when the speed  of player I is bounded by $v$ and $nextToopt(v)$ the function that returns the next to optimal value of the game when the speed  of player I is bounded by $v$. These two functions are decreasing since a strategy for $v$ is always a strategy for $v'\ge v$. Hence, if we have that $opt(v)<nextToopt(v+dv)$ and the strategy at $opt(v)$ is the same as $opt(v+dv)$ it must be that this strategy is optimal in the interval $[v, v+dv]$. By computing $opt(n/1000)$ and $nextToopt(n/1000)$, $n=0,\ldots,1000$ we detect that  the condition stated above is satisfied for $v\in[1/1000, 0.618]$ and for  $v\in[0.619, 0.990]$. To summarize, we have proved
\begin{theorem}(\cite{alpern2006theory}, Chapter 17.1)\label{th:optVSv}
For $v\in[1/1000, 0.618]$ the Rendezvous value of the game is given by $(\ref{eq:exact<})$ and the optimal strategy is plotted on Figure \ref{fig:optVSvpetit} and for  $v\in[0.619, 0.990]$ the Rendezvous value of the game is given by $(\ref{eq:exact>})$ and the optimal strategy is plotted on Figure \ref{fig:optVSvgrand}.
\end{theorem}
The interval in which the Theorem is stated to be true may be enlarged by computing the numerical solutions on a finer mesh, i.e. increasing the number values of $n$ for which we solve the LP.

\section{Solution of the problem with marker held by the slow player}
The marker is held by one of the player and may be dropped off at any given time. Once dropped off, the marker is to be found by the other player when is passes at the location. Once found, the player stops following the original strategy and continues in the same direction until Rendezvous occurs.
We first assume that the marker is held by  player I that moves with the lowest speed bounded by $v\le 1$ and denote $z$ the dropping time. There are $4$ possibilities, $z\in[0,t_1], z\in[t_1, t_2],  z\in[t_2, t_3],  z\in[t_3, t_4]$. Each one leading to a family of linear programs to solve. It occurs that only the first case $z\in[0,t_1]$ is relevant. For all other cases the optimal solutions do not make use of the marker and are given in Section \ref{sec:nomarker}.

The strategy of player I is now given by $(a_0, a_1, a_2, a_3, a_4)$ where $a_i$ indicates whether $v_i$ is positive or negative, i.e. the speed $v_i$ are always assumed positive and the motions are depending on the product $a_i\cdot v_i$. With respect to Section \ref{sec:nomarker} the only novelty is the introduction of $a_0$ see Equation \ref{eq:fbeforet_1} (compare with $(\ref{eq:fnomarker})$ with no marker). 

Agents of player II can find the marker or not. Hence, for each agent we must generate two linear programs each one assuming the agent find the marker or not. Actually, for the first agent Rendezvousing (agent $(o_1,d_1)$)  we do not need to make a difference if the marker is found or not, equations are the same. We use the new variable $k_1$ to indicate that agent $(o_2,d_2)$ finds the marker $k_1=1$ or not ($k_1=0$) in the interval $[0, t_1]$. Again if the marker is found by $(o_2,d_2)$ in the interval $[t_1,t_2]$ the equations do not change. The variable $k_{21}, k_{22}$ indicate that agent $(o_3,d_3)$ finds the marker  in the interval $[0, t_1]$ or $[t_1,t_2]$ respectively. And finally,  the variable $k_{31}, k_{32},k_{33}$ indicate that agent $(o_4,d_4)$ finds the marker  in the interval $[0, t_1]$ or $ [t_1,t_2]$ or $[t_2,t_3]$ respectively.

This leads to the  family of linear programs shown in Equations $(\ref{eq:onemarker})$, $(\ref{eq:onemarker'})$, $(\ref{eq:onemarker''})$ (we denote $\Delta t_i = (t_i-t_{i-1})$ and $t_{iz}$ the time at which the marker is found by $(o_i,b_i)$).

\begin{align}
&\underset{\Delta _{i}}{\text{min}}~~t _{1}+t _{2} +t_{3}+t _{4}  \notag \\
& o_1 + d_1 b_1 t _1 = a_0 vz+a_1 vt_1 \tag{*} \\ \notag \\ 
& k_1(o_2 + d_1 b_2 t_1 + d_2 b_2\Delta t_2 )+\notag \\ & (1-k_1)(o_2 + d_1 b_2 t_{1z} + d_1 b_2(t_1+\Delta t_2-t_{1z})) = a_0 vz+ a_1 vt_1+a_2 vt_2  \notag \\
&(1-k_1)(o_2 + d_1b_2t_{1z}) = (1-k_1)a_0vz \tag{**}\\ \notag \\
& k_{21} k_{22}(o_3 + d_1 b_3 t_1 + d_2 b_3 \Delta t_2 + d_3 b_3 \Delta t_3) + \notag \\
& (1-k_{21})(o_3 + d_1b_3t_{2z} + d_1b_3(t_1+\Delta t_2+\Delta t_3-t_{2z}))+ \label{eq:onemarker} \\
& (1-k_{22})(o_3 + d_1b_3t_1 + d_2b_3t_{2}z + d_2b_3(\Delta t_2+\Delta t_3-t_{2z}))= \notag \\
& a_0 vz+a_1 vt_1+a_2 vt_2 + a_3 vt_3 \notag \\ 
&(1-k_{21})(o_3 + d_1b_3t_{2z})+(1-k_{22})(o_3 + d_1b_3t_1 + d_2b_3t_{2z}) = \tag{***} \\
&(1-k_{21})(1-k_{22})a_0vz \notag
\end{align}

\begin{align}
& (1-k_{31})(1-k_{32})(1-k_{33})(o_4 + d_1 b_4 t_1 + d_2 b_4 \Delta t_2 + d_3 b_4 \Delta t_3 + d_4 b_4 \Delta t_4) + \notag \\
&(1-k_{31})(o_4+d_1b_4t_{3z}+d_1b_4(t_1+\Delta t_2+\Delta t_3+\Delta t_4-t_{3z}))+\notag \\
&(1-k_{32})(o_4+d_1b_4t_1+d_2b_4t_{3z}+d_2b_4(\Delta t_2+\Delta t_3+\Delta t_4-t_{3z})) + \label{eq:onemarker'} \\
&(1-k_{33})(o_4+d_1b_4t_1+d_2b_4\Delta t_2+d_3b_4t_{3z}+d_3b_4(\Delta t_3+\Delta t_4-t_{3z}))=\notag \\
& a_o vz+a_1 vt_1+ a_2 vt_2 + a_3 vt_3 + a_4 vt_4  \notag \\ \notag
&(1-k_{31})(o_4+d_1b_4t_{3z})+(1-k_{32})(o_4+d_1b_4t_1+d_2b_4t_{3z}) + \notag \\
& (1-k_{33})(o_4+d_1b_4t_1+d_2b_4\Delta t_2+d_3b_4t_{3z})= \tag{****} \\
&a_0vz(1-k_{31})(1-k_{32})(1-k_{33}) \notag \\ \notag
\end{align}

In this set of equations, the first one is the minimization  problem to be solved. Notice that we minimize the sum of the Rendezvous time while the number given in the Introduction and results are the average of the Rendezvous times. There are four sets of equations, $(*), (**), (***), (****)$. Equation $(*)$ is the constraint that player I Rendezvous with agent $(o_1,b_1)$ at time $t_1$. Agent $(o_1,b_1)$ may find the marker before time $t_1$. However, in this case the optimal solution continues in the same direction, i.e. the equation would be 
\begin{equation*}
k( o_1 +d_1 b_1 t _1)+(1-k)(o_1 + d_1*t_{0z}d_1 b_1 (t _1-t_{0z}) = a_0 vz+a_1 vt_1, 
\end{equation*}
where $k=1$ if player II does not find the marker and $k=0$  else and $t_{oz}$ is the time at which player II finds the marker. This equation reduces to $(*)$.  In $(*)$ if player finds the marker in the interval $[t_1,t_2]$ the optimal strategy is to continue the same direction and the marker is useless, and so on for $(***)$ and $(****)$  if the marker is found in the interval $[t_2,t_3]$ and $[t_3,t_4]$ respectively.

The three remaining sets of equations $(**), (***), (****)$ are composed of two equations. The first one accounts for the Rendezvous of agent $(o_i,b_i)$ with player I and the second one is valid only if the marker is used ($k_1, k_{21}, k_{22}, k_{31}, k_{32}, k_{33}$ equal $1$) and define the times when the marker is found $t_{1z}, t_{2z}, t_{3z}$.

The next set of equations is composed of the speed constraints. The variable $vz$ is the product of the speed of player I and the time $z$ of dropping the marker, this product is bounded by $v\cdot z$ since the speed of player I is bounded by $v$. In the results we observe that the speed of player I is $v$ (maximal) or $0$ but we obtain no solution with $v$ in between.
\begin{equation}\label{eq:onemarker''}
\begin{split}
& 0 \le vz \le v\cdot z \\
& 0\le vt_1 \le v \cdot (t_1-z)\\
& 0\le vt_2 \le v\cdot \Delta t_2\\
& 0\le vt_3 \le v\cdot  \Delta t_3\\
& 0\le vt_4 \le v\cdot  \Delta t_4
\end{split}
\end{equation}

The family of linear programs is generated by assigning values to the parameters of equations $(\ref{eq:onemarker})$, $(\ref{eq:onemarker'})$, $(\ref{eq:onemarker''})$. These values must satisfy the constraints (the constraints $t_i\ge 0$ are included in all linear programs)
\begin{align}
& a_i,b_i, o_i, d_i\in \{0,1\} \notag \sum o_{i}=0, \sum d_{i}=0, ~o_i=o_j \Rightarrow d_i\not = d_j, \label{eq:onemarker'''}\\
& t_1 \ge 0, t_2 \ge 0, t_3 \ge 0, t_4 \ge 0\notag.
\end{align}

\begin{figure}[h]
\begin{center}
\includegraphics[scale=0.4]{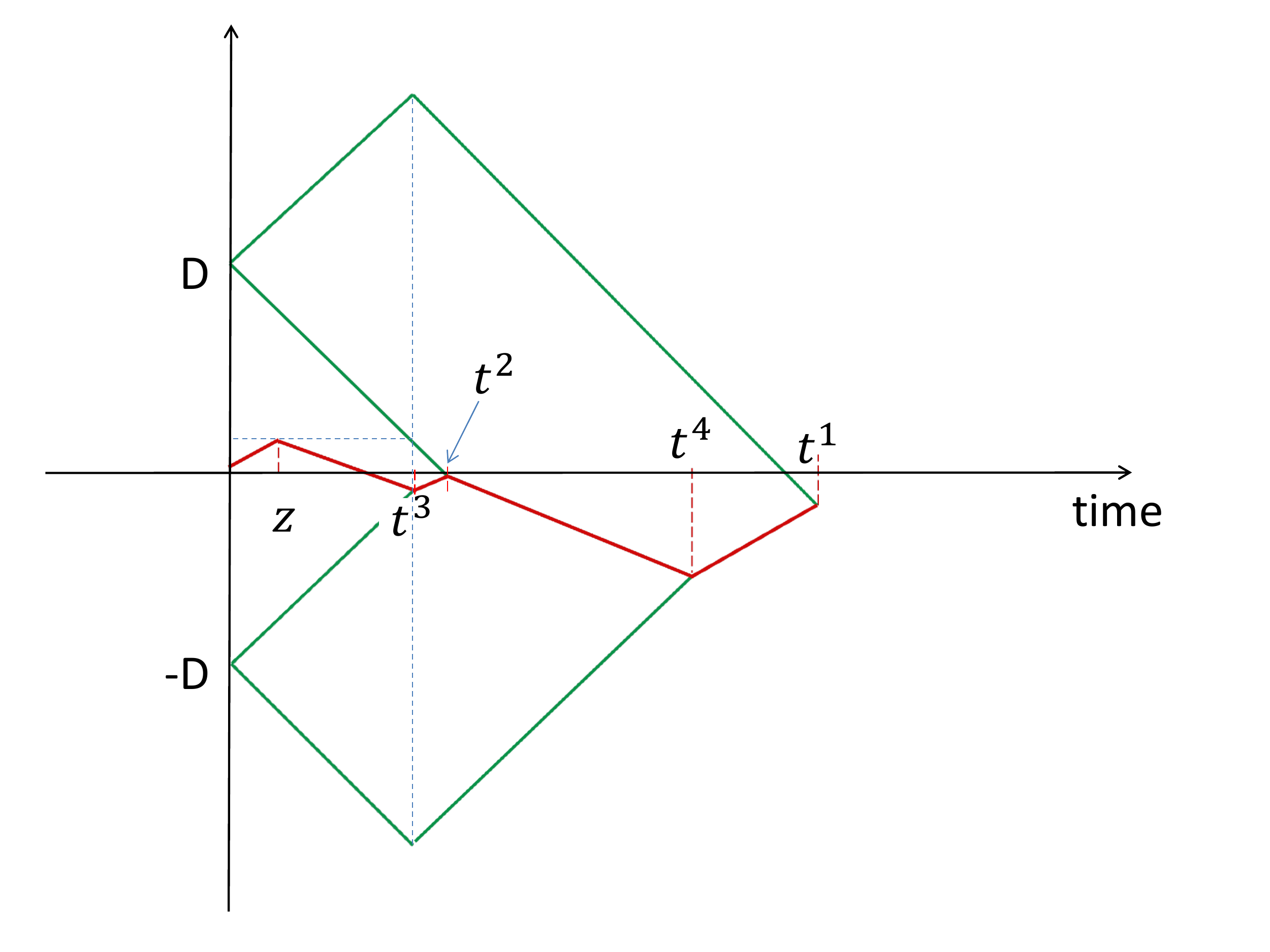}
\end{center}\caption{Optimal strategy when player I holds the marker and is the slower player.}\label{fig:optstratmarkerI}
\end{figure}

The families of linear programs are solved for maximal speed of player I ranging from $0$ to $1$ with a step size of $1/1000$, i.e.  optimal solutions $opt(v)$  are computed for $v=n/1000$, $n=0,1,\ldots,1000$.  The result is that the same strategy is used, see Figure \ref{fig:optstratmarkerI}. The speed of player I is maximal along the trajectory and the best solution is obtained for $z\in[0,t_1]$. With respect to the notation of Corollary \ref{cor:optstratv} the speeds of player I are $(v_0=-v, v_1=v,v_2=-v,v_3=v)$. The marker reduces the Rendezvous value even when the speed of player I is very slow.

The Rendezvous times are given by
\begin{equation}\label{eq:optRwithmarkerI}
\begin{split} 
z&=\frac{1}{v+3},~ t_1=\frac{3}{v+3},~  t_2=\frac{5v+3}{(v+1)(v+3)},\\
t_3&=\frac{7v^2+12v+9}{(v+1)^2(v+3)},~ t_4=\frac{9v^3+27v^2+35v+9}{(v+1)^3(v+3)},\\ \\
R&=\frac{24v^3+68v^2+76v+24}{(v+1)^3(v+3)}.
\end{split}
\end{equation}

\begin{figure}[h]
\begin{center}
\includegraphics[scale=0.6]{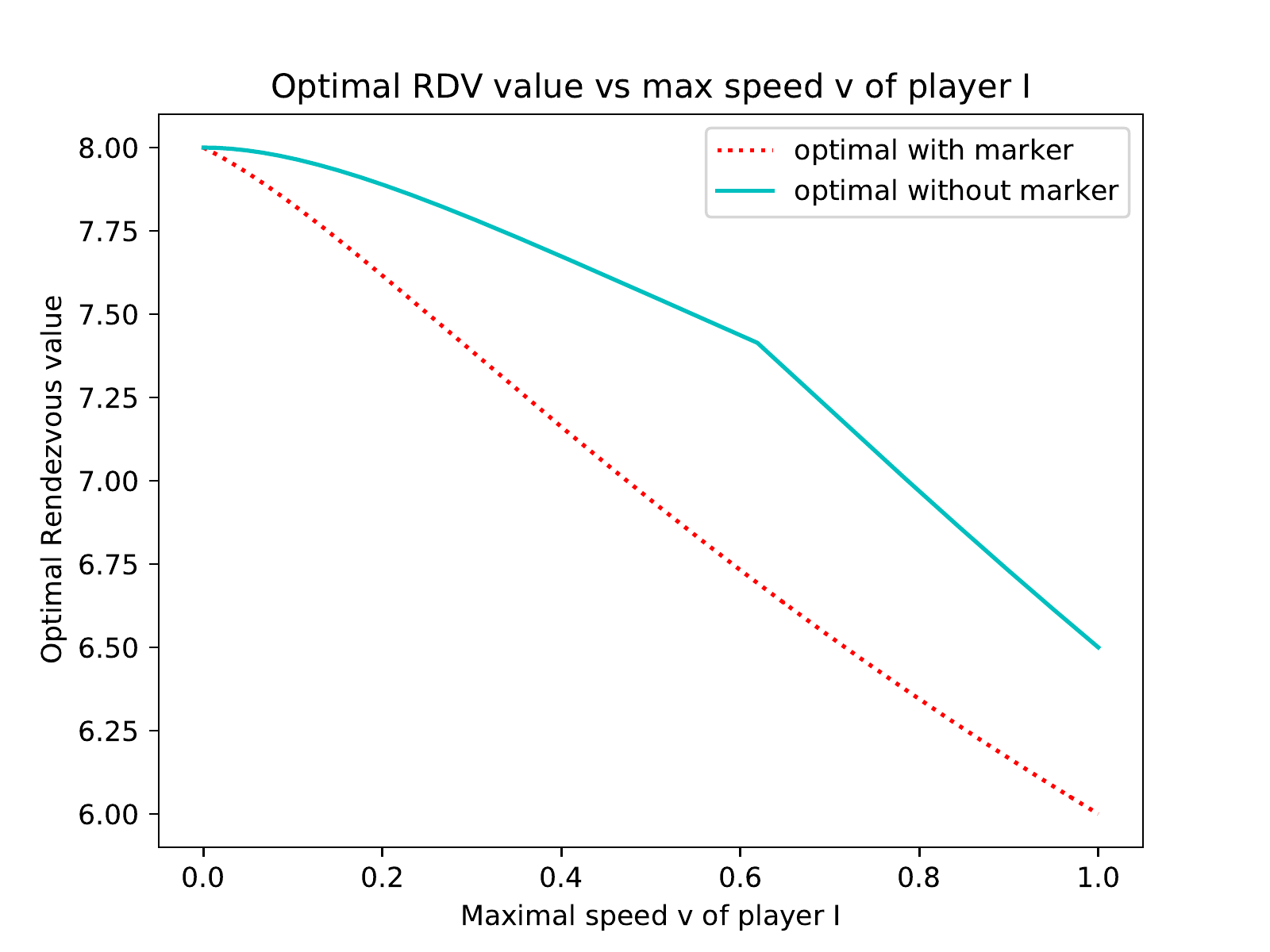}
\end{center}\caption{Optimal solutions with $(\ref{eq:optRwithmarkerI})$  and without marker. The slow player holds the marker.}\label{fig:optVSvmarker}
\end{figure}

The optimal solution function $opt(v)$ is decreasing because a strategy for $v$ is a strategy for $v'\ge v$ as well. Hence, if the value of the next to optimal strategy for speed $v+dv$, $nextToopt(v+dv)$ is larger than $opt(v)$ the strategy that leads to $opt(v)$ is optimal for speeds in $[v,v+dv]$. With our mesh size of $1/1000$ we numerically observe that this occurs since $v\ge 17/1000$. Hence we have a computer assisted proof summarized in the following Theorem.

\begin{theorem}
For $v\in[17/1000, 1]$ the Rendezvous value of the game is given by $(\ref{eq:optRwithmarkerI})$. The optimal strategy is plotted on Figure \ref{fig:optstratmarkerI} and the optimal Rendezvous values on Figure \ref{fig:optVSvmarker}.
\end{theorem}
The interval on which the Theorem is stated to be true may be enlarged by computing the numerical solutions on a finer mesh. It is relevant to point out that for the optimal strategy the time at which the marker is found is $t_1$, i.e. the same time as the first Rendezvous occurs.

\section{Solution of the problem with marker held by the fast player}
The family of linear programs to be solved when the marker is held by the fast player (I) is very similar to the one defined by Equations $(\ref{eq:onemarker})$, $(\ref{eq:onemarker'})$, $(\ref{eq:onemarker''})$. The changes are that the coefficients $a_i$ are no longer multiplied by the maximal speed $v$ as are now the coefficients $d_i$. The system of equations is not reproduced here to save some space.

\begin{figure}[h]
\begin{center}
\includegraphics[scale=0.6]{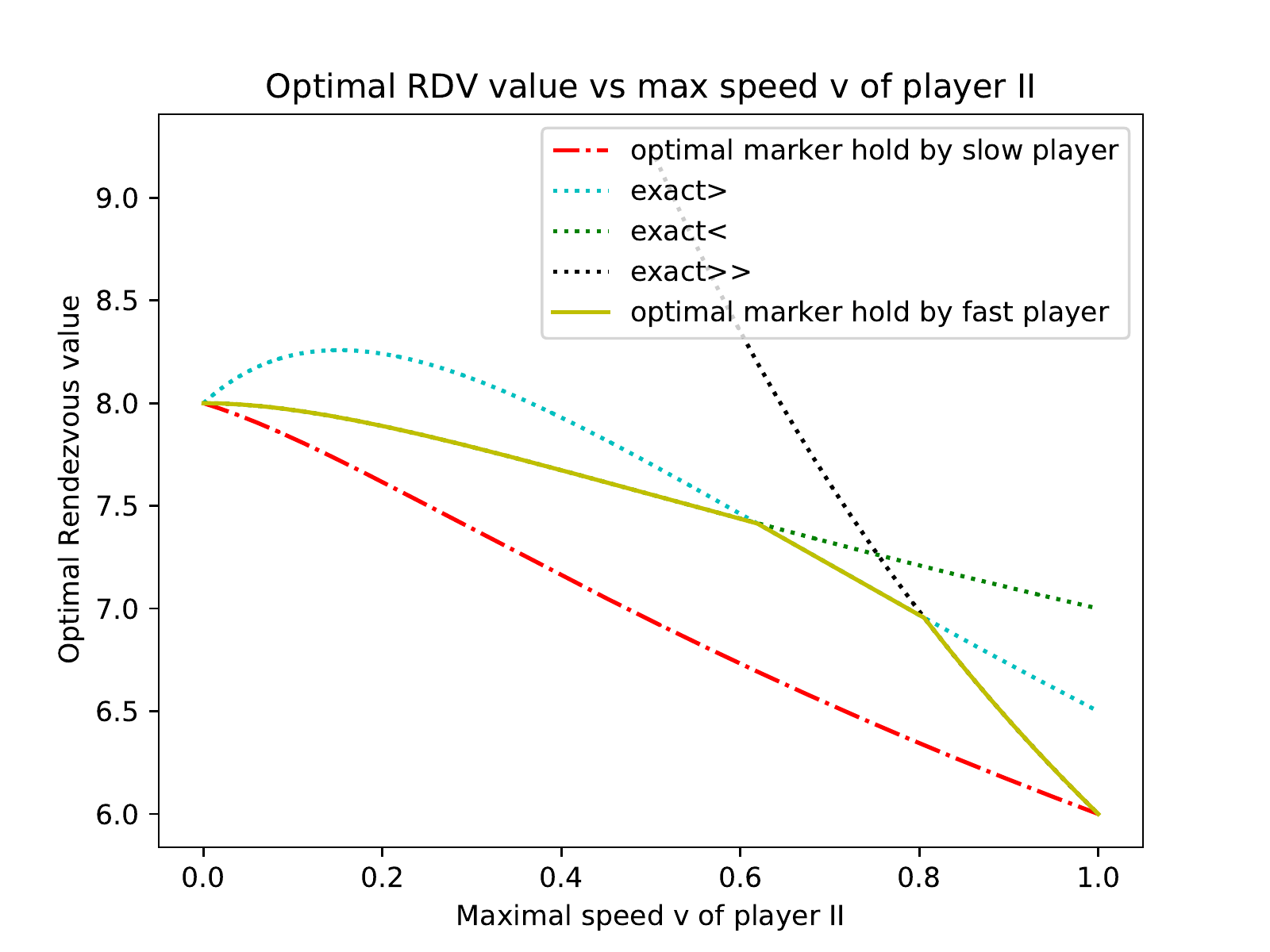}
\end{center}\caption{Optimal solutions with marker, i.e $(\ref{eq:exact<})$ for $v\le (\sqrt{5}-1)/2$, $(\ref{eq:exact>})$ for $v\ge (\sqrt{5}-1)/2$, $(\ref{eq:optRwithmarker I-II})$ for $v > 0.805$,  and without marker. The fast player holds the marker.}\label{fig:optVSvmarker I-II}
\end{figure}

The results are plotted on Figure \ref{fig:optVSvmarker I-II}. We observe that for speeds slower than $v \approx 0.805$ the marker is not useful and the optimal solution is given by the optimal solutions without marker stated in Theorem \ref{th:optVSv}. For speeds faster than $v\approx 0.805$ the marker start to be useful and the strategy is similar to the optimal one when the marker is held by the slow player, see Figure \ref{fig:optstratmarkerI}. When the fast player holds the marker we obtain that the Rendezvous times are given by:
\begin{equation}\label{eq:optRwithmarker I-II}
\begin{split} 
z&=\frac{1}{3v+1},~ t_1=\frac{3}{3v+1},~  t_2=\frac{3v+5}{(v+1)(3v+1)},\\
t_3&=\frac{9v+5}{(v+1)(3v+1)},~ t_4=\frac{3v+7}{(v+1)^2},\\ \\
R&=\frac{24v^2+52v+20}{(v+1)^2(3v+1)}.
\end{split}
\end{equation}

The optimal Rendezvous value is decreasing with the maximal speed $v$ because a stragey for maximal speed $v$ is a strategy for maximal speed $v'\ge v$. Moreover, if we denote $opt(v)$ the optimal Rendezvous value for maximal speed $v$, and $nextToopt(v)$ the next to optimal Rendezvous value, it follows that if $nextToopt(v+dv)\ge opt(v)$ and the strategy leading to $opt(v)$ and $opt(v+dv)$ is the same $\Longrightarrow$ the strategy is optimal on the entire interval $[v,v+dv]$. By numerical computation and using the two stated observations we obtain the following Theorem.

\begin{figure}[h]
\begin{center}
\includegraphics[scale=0.4]{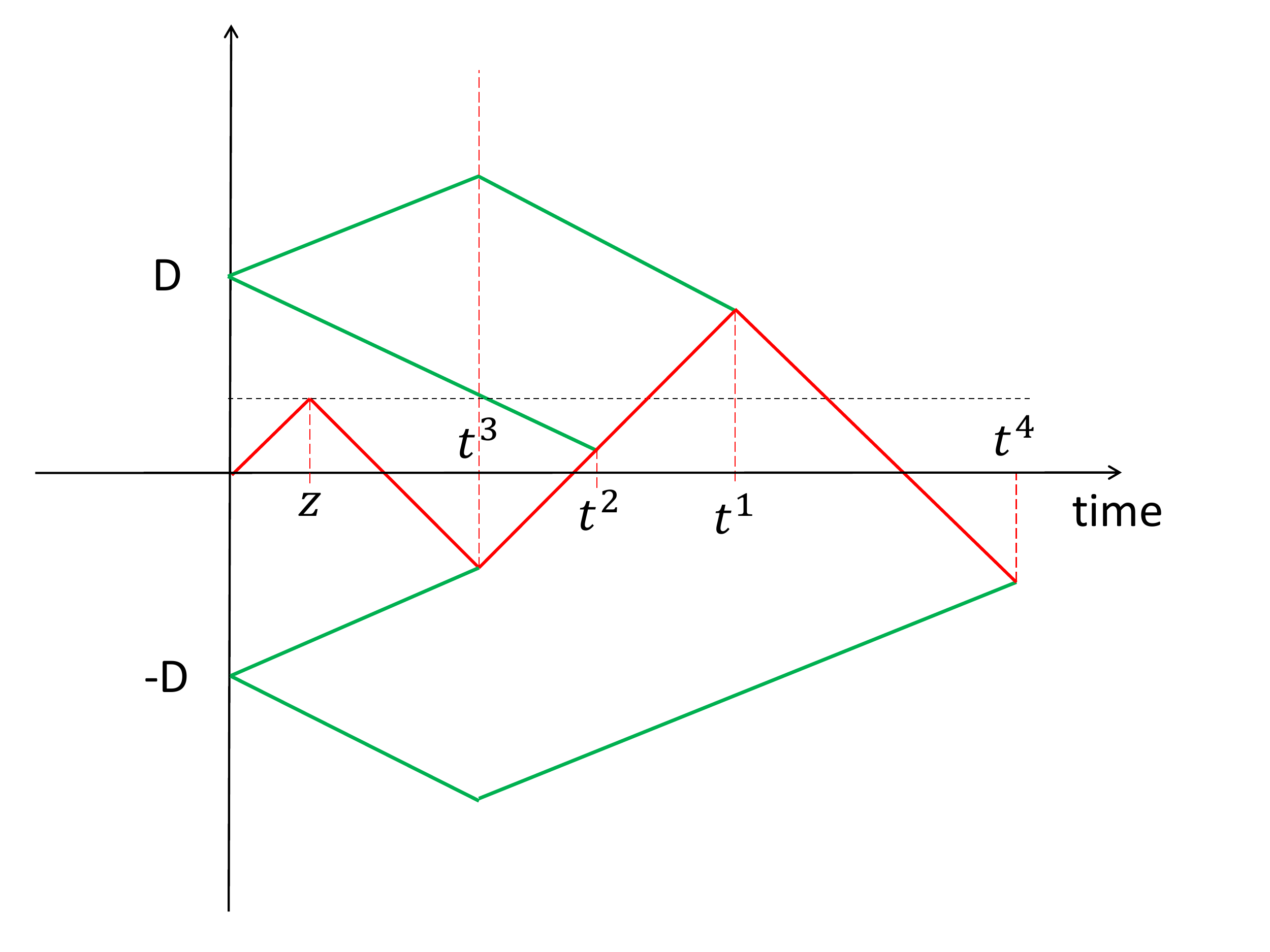}
\end{center}\caption{Optimal solutions with marker, i.e $(\ref{eq:optRwithmarker I-II})$ for $v > 0.805$, for lower speeds the optimal solutions are the ones without marker (Theorem \ref{th:optVSv}). The fast player holds the marker.}\label{fig:optVSvgrand I-II}
\end{figure}

\begin{theorem}\label{th:optVSvmarker I-II}
For $v\in[1/1000, 0.618]$ the Rendezvous value of the game is given by $(\ref{eq:exact<})$ and the optimal strategy is plotted on Figure \ref{fig:optVSvpetit} and for  $v\in[0.619, 0.805]$ the Rendezvous value of the game is given by $(\ref{eq:exact>})$ and the optimal strategy is plotted on Figure \ref{fig:optVSvgrand} and for  $v\in[0.807, 0.966]$ the Rendezvous value of the game is given by $(\ref{eq:optRwithmarker I-II})$, the Rendezvous values are plotted on Figure \ref{fig:optVSvmarker I-II}  and the optimal strategy is plotted on Figure \ref{fig:optVSvgrand I-II}.
\end{theorem}

It is relevant to point out the difference between the optimal strategies when the marker is held by the slow (Figure \ref{fig:optstratmarkerI}) or fast player (Figure \ref{fig:optVSvgrand I-II}). After Rendezvous time $t_2$ on Figure \ref{fig:optstratmarkerI} the slow player (who holds the marker) turns while on Figure \ref{fig:optVSvgrand I-II} the fast player (who holds the marker) continues on his way. The transition from the two strategies is `continuous' in the sense that when the speeds are equal at time $t_2$ the two remaining agents to be found are at equal distances from player I. Hence, both strategies are optimal (turning or continuing). 


\noindent{\bf Acknowledgment:}
Data sharing not applicable to this article as no datasets were generated or analysed during the current study. On behalf of all authors, the corresponding author states that there is no conflict of interest. This work has not been funded by any special grant
\newpage
\bibliography{biblioRendezVous}

\begin{thebibliography}{10}

\bibitem{doi:10.1137/S0363012993249195}
Steve Alpern.
\newblock The rendezvous search problem.
\newblock {\em SIAM Journal on Control and Optimization}, 33(3):673--683, 1995.

\bibitem{alpern2013ten}
Steve Alpern.
\newblock Ten open problems in rendezvous search.
\newblock In {\em Search Theory}, pages 223--230. Springer, 2013.

\bibitem{alpern1999asymmetric}
Steve Alpern and Anatole Beck.
\newblock Asymmetric rendezvous on the line is a double linear search problem.
\newblock {\em Mathematics of Operations Research}, 24(3):604--618, 1999.

\bibitem{alpernbeck1999}
Steve Alpern and Anatole Beck.
\newblock Rendezvous search on the line with limited resources: Maximizing the
  probability of meeting.
\newblock {\em Operations Research}, 47(6):849--861, 1999.

\bibitem{alpernbeck2000}
Steve Alpern and Anatole Beck.
\newblock Pure strategy asymmetric rendezvous on the line with an unknown
  initial distance.
\newblock {\em Operations Research}, 48(3):498--501, 2000.

\bibitem{alpern2013search}
Steve Alpern, Robbert Fokkink, L~Gasieniec, Roy Lindelauf, and VS~Subrahmanian.
\newblock {\em Search theory}.
\newblock Springer, 2013.

\bibitem{alperngal1995}
Steve Alpern and Shmuel Gal.
\newblock Rendezvous search on the line with distinguishable players.
\newblock {\em SIAM Journal on Control and Optimization}, 33(4):1270--1276,
  1995.

\bibitem{alpern2006theory}
Steve Alpern and Shmuel Gal.
\newblock {\em The theory of search games and rendezvous}, volume~55.
\newblock Springer Science \& Business Media, 2006.

\bibitem{andersonessegaier1995}
Edward~J. Anderson and Skander Essegaier.
\newblock Rendezvous search on the line with indistinguishable players.
\newblock {\em SIAM Journal on Control and Optimization}, 33(6):1637--1642,
  1995.

\bibitem{baston1999}
Vic Baston.
\newblock Note: Two rendezvous search problems on the line.
\newblock {\em Naval Research Logistics (NRL)}, 46(3):335--340, 1999.

\bibitem{bastongal1998}
Vic Baston and Shmuel Gal.
\newblock Rendezvous on the line when the players' initial distance is given by
  an unknown probability distribution.
\newblock {\em SIAM Journal on Control and Optimization}, 36(6):1880--1889,
  1998.

\bibitem{baston2001rendezvous}
Vic Baston and Shmuel Gal.
\newblock Rendezvous search when marks are left at the starting points.
\newblock {\em Naval Research Logistics (NRL)}, 48(8):722--731, 2001.

\bibitem{doi:10.1002/net.21504}
Vic Baston and Kensaku Kikuta.
\newblock Search games on networks with travelling and search costs and with
  arbitrary searcher starting points.
\newblock {\em Networks}, 62(1):72--79, 2013.

\bibitem{bastonkikuta}
Vic Baston and Kensaku Kikuta.
\newblock {Search games on a network with travelling and search costs}.
\newblock {\em International Journal of Game Theory}, 44(2):347--365, May 2015.

\bibitem{baston2019search}
Vic Baston and Kensaku Kikuta.
\newblock A search problem on a bipartite network.
\newblock {\em European Journal of Operational Research}, 277(1):227--237,
  2019.

\bibitem{chrobak2015group}
Marek Chrobak, Leszek G{\k{a}}sieniec, Thomas Gorry, and Russell Martin.
\newblock Group search on the line.
\newblock In {\em International Conference on Current Trends in Theory and
  Practice of Informatics}, pages 164--176. Springer, 2015.

\bibitem{czyzowicz2008power}
Jurek Czyzowicz, Stefan Dobrev, Evangelos Kranakis, and Danny Krizanc.
\newblock The power of tokens: rendezvous and symmetry detection for two mobile
  agents in a ring.
\newblock In {\em International Conference on Current Trends in Theory and
  Practice of Computer Science}, pages 234--246. Springer, 2008.

\bibitem{czyzowicz2018linear}
Jurek Czyzowicz, Ryan Killick, and Evangelos Kranakis.
\newblock Linear rendezvous with asymmetric clocks.
\newblock In {\em 22nd International Conference on Principles of Distributed
  Systems (OPODIS 2018)}. Schloss Dagstuhl-Leibniz-Zentrum fuer Informatik,
  2018.

\bibitem{das2019gathering}
Shantanu Das, Riccardo Focardi, Flaminia~L Luccio, Euripides Markou, and Marco
  Squarcina.
\newblock Gathering of robots in a ring with mobile faults.
\newblock {\em Theoretical Computer Science}, 764:42--60, 2019.

\bibitem{das2015mobile}
Shantanu Das, Flaminia~L Luccio, and Euripides Markou.
\newblock Mobile agents rendezvous in spite of a malicious agent.
\newblock In {\em International Symposium on Algorithms and Experiments for
  Wireless Sensor Networks}, pages 211--224. Springer, 2015.

\bibitem{das2008rendezvous}
Shantanu Das, Mat{\'u}{\v{s}} Mihal{\'a}k, Rastislav {\v{S}}r{\'a}mek, Elias
  Vicari, and Peter Widmayer.
\newblock Rendezvous of mobile agents when tokens fail anytime.
\newblock In {\em International Conference On Principles Of Distributed
  Systems}, pages 463--480. Springer, 2008.

\bibitem{di2017optimal}
Gabriele Di~Stefano and Alfredo Navarra.
\newblock Optimal gathering of oblivious robots in anonymous graphs and its
  application on trees and rings.
\newblock {\em Distributed Computing}, 30(2):75--86, 2017.

\bibitem{flocchini2004mobile}
Paola Flocchini, Evangelos Kranakis, Danny Krizanc, Flaminia~L Luccio, Nicola
  Santoro, and Cindy Sawchuk.
\newblock Mobile agents rendezvous when tokens fail.
\newblock In {\em International Colloquium on Structural Information and
  Communication Complexity}, pages 161--172. Springer, 2004.

\bibitem{flocchini2004multiple}
Paola Flocchini, Evangelos Kranakis, Danny Krizanc, Nicola Santoro, and Cindy
  Sawchuk.
\newblock Multiple mobile agent rendezvous in a ring.
\newblock In {\em Latin American Symposium on Theoretical Informatics}, pages
  599--608. Springer, 2004.

\bibitem{han2008}
Qiaoming Han, Donglei Du, Juan Vera, and Luis~F. Zuluaga.
\newblock Improved bounds for the symmetric rendezvous value on the line.
\newblock {\em Operations Research}, 56(3):772--782, 2008.

\bibitem{hohzaki2016search}
Ryusuke Hohzaki.
\newblock Search games: Literature and survey.
\newblock {\em Journal of the Operations Research Society of Japan},
  59(1):1--34, 2016.

\bibitem{howard1999}
J.~V. Howard.
\newblock Rendezvous search on the interval and the circle.
\newblock {\em Operations Research}, 47(4):550--558, 1999.

\bibitem{kranakis2010mobile}
Evangelos Kranakis, Danny Krizanc, and Euripides Markou.
\newblock The mobile agent rendezvous problem in the ring.
\newblock {\em Synthesis lectures on distributed computing theory},
  1(1):1--122, 2010.

\bibitem{kranakis2003mobile}
Evangelos Kranakis, Nicola Santoro, Cindy Sawchuk, and Danny Krizanc.
\newblock Mobile agent rendezvous in a ring.
\newblock In {\em 23rd International Conference on Distributed Computing
  Systems, 2003. Proceedings.}, pages 592--599. IEEE, 2003.

\bibitem{leone2018rendezvous}
Pierre Leone and Steve Alpern.
\newblock Rendezvous search with markers that can be dropped at chosen times.
\newblock {\em Naval Research Logistics (NRL)}, 65(6-7):449--461, 2018.

\bibitem{leonealpernNRL2018}
Pierre Leone and Steve Alpern.
\newblock Rendezvous search with markers that can be dropped at chosen times.
\newblock {\em Naval Research Logistics (NRL)}, 65(6-7), 2018.

\bibitem{DBLP:journals/eor/Lidbetter20}
Thomas Lidbetter.
\newblock Search and rescue in the face of uncertain threats.
\newblock {\em Eur. J. Oper. Res.}, 285(3):1153--1160, 2020.

\bibitem{ozsoyeller2013}
D.~Ozsoyeller, A.~Beveridge, and V.~Isler.
\newblock Symmetric rendezvous search on the line with an unknown initial
  distance.
\newblock {\em IEEE Transactions on Robotics}, 29(6):1366--1379, Dec 2013.

\bibitem{pelc2019deterministic}
Andrzej Pelc.
\newblock Deterministic rendezvous algorithms.
\newblock In {\em Distributed Computing by Mobile Entities}, pages 423--454.
  Springer, 2019.

\bibitem{stachowiak2009asynchronous}
Grzegorz Stachowiak.
\newblock Asynchronous deterministic rendezvous on the line.
\newblock In {\em International Conference on Current Trends in Theory and
  Practice of Computer Science}, pages 497--508. Springer, 2009.

\bibitem{Uthaisombut2005SymmetricRS}
Patchrawat Uthaisombut.
\newblock Symmetric rendezvous search on the line using move patterns with
  different lengths.
\newblock Citeseer, 2006.

\end{thebibliography}
\bibliographystyle{plain}

\end{document}